\documentclass[letterpaper,final]{appolb}

 \usepackage{latexsym,bm,amsmath,amssymb,amsfonts,slashed}
 \usepackage{epsfig,graphics,graphicx}
 \usepackage{cite}

 \long\def\comment#1{ }
 \newcommand{\eqn}[1]{Eq.~\eqref{#1}}
 \newcommand{\beq}{\begin{equation}}
 \newcommand{\eeq}{\end{equation}}
 \newcommand{\nn}{\nonumber\\}
 \newcommand{\dif}{{\rm d}}
 \newcommand{\rmd}{{\rm d}}
 \newcommand{\rme}{{\rm e}}
 \newcommand{\rmi}{{\rm i}}
 \newcommand{\rmtr}{{\rm tr}}
 
 \newcommand{\del}{\partial}
 \newcommand{\lan}{\langle}
 \newcommand{\ran}{\rangle}

 \newcommand{\order}[1]{\mcal{O}{(#1)}}
 \newcommand{\mcal}{\mathcal}
 \newcommand{\wt}{\widetilde}
 \newcommand{\bmk}{\bm{k}}
 \newcommand{\bmp}{\bm{p}}
 
 \newcommand{\bmx}{\bm{x}}
 
 \newcommand{\bmu}{\bm{u}}
 \newcommand{\bmv}{\bm{v}}
 \newcommand{\bmz}{\bm{z}}
 \newcommand{\bmr}{\bm{r}}
 
 \newcommand{\abar}{\bar{\alpha}}
 \newcommand{\Lam}{\Lambda_{{\rm QCD}}}

\begin{document}

\title{THE COLOR GLASS CONDENSATE AND SOME APPLICATIONS\thanks{Presented at the Workshop ``Excited QCD 2012'', Peniche, Portugal, May 6-12, 2012.}}
\author{D.N.~Triantafyllopoulos
\address{ECT*, Strada delle Tabarelle 286, I-38123, Villazzano (TN), Italy}}
\maketitle


\begin{abstract}
We give an elementary discussion of parton saturation and its description by the effective theory of the Color Glass Condensate. We report on progress in calculating multi-gluon correlators. The latter are necessary for many phenomenological applications, upon some of which we briefly touch. 
\end{abstract}

\section{From partons to the Color Glass Condensate}
\label{sec:partons}

A hadron is a complex object in its rest frame as valence quarks are accompanied by a sea of hadronic and vacuum fluctuations. Both types of such virtual fluctuations are of non-perturbative nature and have the same lifetime $\Delta t_{\rm RF} \sim 1/\Lam$, since $\Lam$ is the only available energy scale. The picture changes in the infinite momentum frame; the lifetime of the vacuum ones remains the same (the vacuum is Lorentz invariant), but the hadronic ones are time-dilated, that is $\Delta t_{\rm IMF} \sim \gamma/\Lam$ with $\gamma \gg 1$, and the two types are disentangled \cite{Iancu:2012xa}. Probing the hadronic fluctuations is most easily done by DIS of a lepton off the hadron. The collision time  is estimated as $\Delta t_{\rm coll} \sim 2 x P/Q^2$. Thus, quarks with transverse momenta $k_{\perp} \ll Q$, will be ``seen'' by the virtual photon, since $\Delta t_{\rm fluct} \sim 2 x P/k_{\perp}^2 \gtrsim \Delta t_{\rm coll}$. 

The density of partons depends on their momenta and is determined by the QCD dynamics. The probability for a parton in representation $R$ with momenta $(\bm{0}_{\perp},p_z)$ to emit a soft gluon with $(\bm{k}_{\perp},k_z=x p_z)$ at small-$x$ is 
 \beq\label{prob}
 \dif P = C_R\,\frac{\alpha_s(k_{\perp}^2)}{\pi^2}\,
 \frac{\dif^2 \bmk_{\perp}}{k_{\perp}^2}\,\frac{\dif x}{x}.
 \eeq
The emission of soft and collinear gluons is favored, since it goes with a large logarithm which may even overcome the smallness of the QCD coupling. Such radiative processes need to be resummed to all orders and lead to the DGLAP and BFKL evolution equations, in $\ln Q^2$ and $\ln 1/x$ respectively, and the one to be used depends on the kinematics. The last factor in \eqn{prob} is present because the emitted parton is a gluon, hence the small-$x$ tail of the wavefunction is dominated by gluons. For a BFKL cascade generated by a single valence quark one finds the unintegrated gluon distribution, i.e.~the number of gluons per unit rapidity $Y=\ln 1/x$ and for a given $k_{\perp}$, to increase exponentially with $Y$.

\begin{figure}[t]
 \begin{center}
 \begin{minipage}[b]{0.48\textwidth}
 \begin{center}
 \includegraphics[scale=0.45]{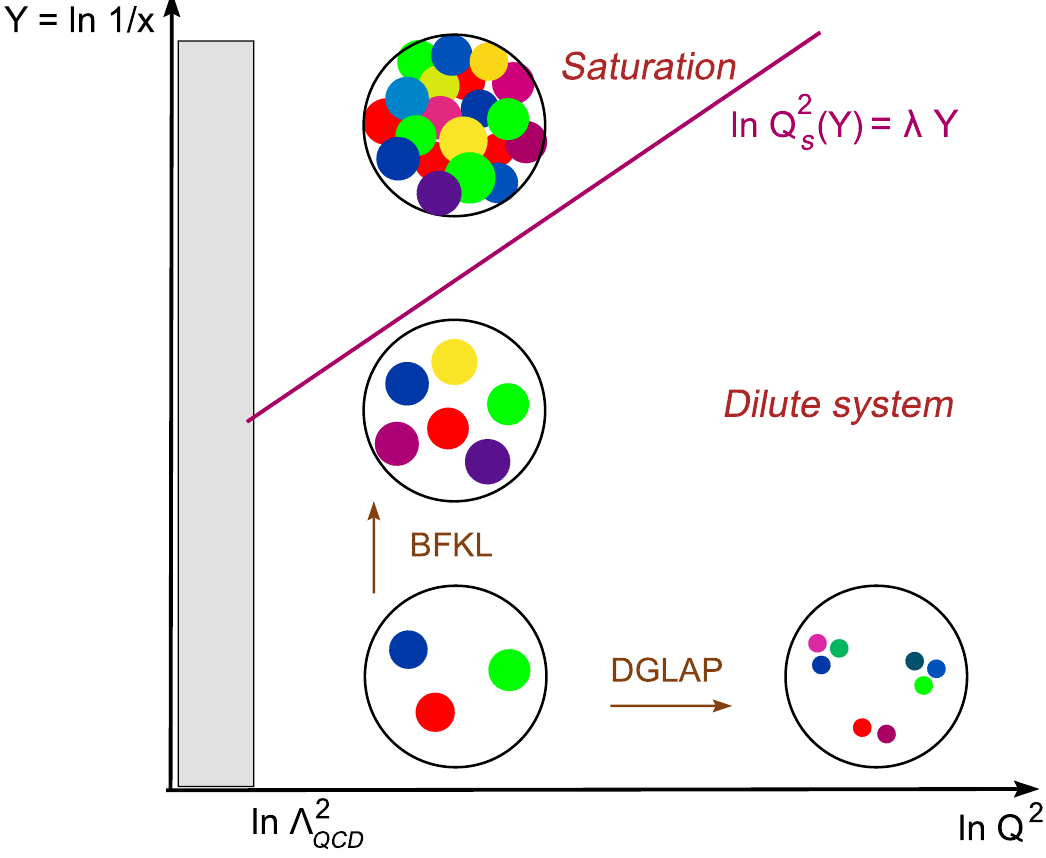}\\{\footnotesize (a)}
 \end{center}
 \end{minipage}
 \quad
 \begin{minipage}[b]{0.48\textwidth}
 \begin{center}
 \includegraphics[scale=0.22]{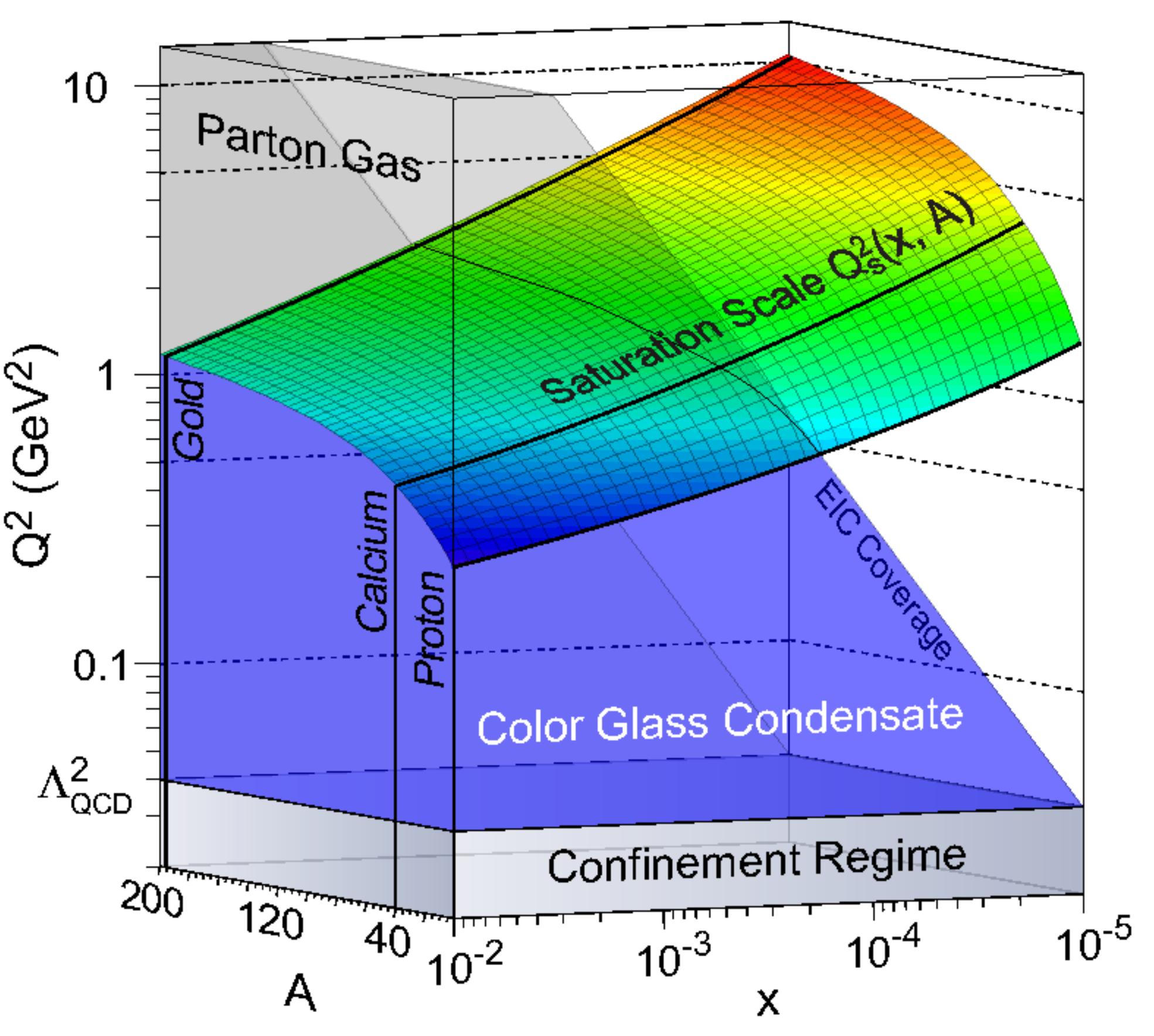}\\{\footnotesize (b)}
 \end{center}
 \end{minipage}
 \caption{(a) The partonic phase diagram and (b) the saturation momentum.
 \label{fig:sat}}
 \end{center}
 \vspace*{-0.5cm}
 \end{figure}

DGLAP and BFKL evolutions are linear and emissions from the source partons take place independently but they lead to totally different outcomes, cf.~Fig.~\ref{fig:sat}.a. Increasing $Q^2$, DGLAP produces more partons (for $x \lesssim 0.1$), but they occupy a transverse area $1/Q^2$ and thus the partonic system becomes more dilute. It is self-consistent in the sense that linear dynamics remains valid in the course of evolution. On the contrary, increasing $\ln 1/x$, BFKL produces partons typically of the same size which eventually overlap. Beyond that point the source partons emit coherently, the BFKL equation becomes inadequate and has to be supplemented by non-linear terms which lead to parton saturation. A simple criterion to define saturation is via the gluon occupation number, the number of gluons at a given $x$ multiplied by the area each gluon fills up and divided by the transverse hadron size. We define the saturation momentum as the line along which this occupation factor is constant and of order $1/\alpha_s$, namely $xg(x,Q_s^2(x))/[Q_s^2(x) R^2] \sim 1/\alpha_s$, with $R$ the hadron size and $xg$ the integrated gluon distribution. Roughly, $Q_s^2$ is increasing as a small power of $1/x$ \cite{Triantafyllopoulos:2002nz,Beuf:2010aw,Avsar:2011ds}, reflecting the need to stop the exponential growth in $Y$ discussed earlier and for a large nucleus is proportional to $A^{1/3}$, since $xg$ grows with $A$ and $R_A$ with $A^{1/3}$. Cf.~Fig.~\ref{fig:sat}.b.

A modern effective theory encompassing parton saturation is the Color Glass Condensate (CGC) \cite{Iancu:2012xa}. Color since we are dealing with QCD, glass since there is a separation in times scales--source partons live longer than emitted ones--, and condensate since occupation numbers can be as high as $1/\alpha_s$. It describes the hadron in terms of a classical field $\mcal{A}$ which reaches values of order $1/g$. Parton distributions are replaced with a classical probability distribution $W_Y[\mcal{A}]$ encoding higher-point correlations. The evolution of this functional is quantum and obtained by summing all orders in $\abar Y$, with $\abar=\alpha_s N_c/\pi$, like in BFKL, and all orders in the field $\mcal{A}$. One has \cite{JalilianMarian:1997gr,Weigert:2000gi,Iancu:2000hn}
 \beq\label{JIMWLK}
 \lan \hat{\mcal{O}}[\mcal{A}] \ran_Y = 
 \int \mcal{D} \mcal{A} \,W_Y[\mcal{A}]\, \hat{\mcal{O}}[\mcal{A}]
 \quad
 \mathrm{with}
 \quad
 \frac{\del W_Y[\mcal{A}]}{\del Y} = H\, W_Y[\mcal{A}],
 \eeq
with $\hat{\mcal{O}}[\mcal{A}]$ a generic observable and $H$ the JIMWLK Hamiltonian.

\section{Particle production in $pA$ collisons}
\label{sec:pA}

\begin{figure}[t]
 \begin{center}
 \begin{minipage}[b]{0.325\textwidth}
 \begin{center}
 \includegraphics[scale=0.45]{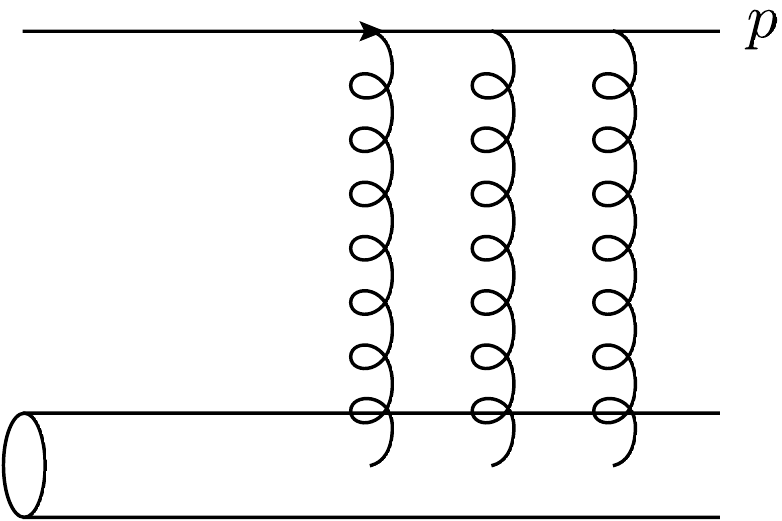}\\{\footnotesize (a)}
 \end{center}
 \end{minipage}
 \begin{minipage}[b]{0.325\textwidth}
 \begin{center}
 \includegraphics[scale=0.45]{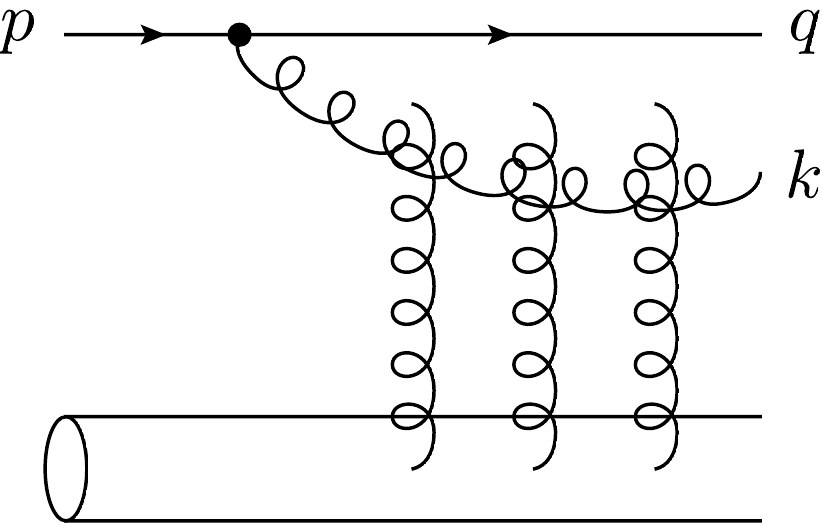}\\{\footnotesize (b)}
 \end{center}
 \end{minipage}
 \begin{minipage}[b]{0.325\textwidth}
 \begin{center}
 \includegraphics[scale=0.45]{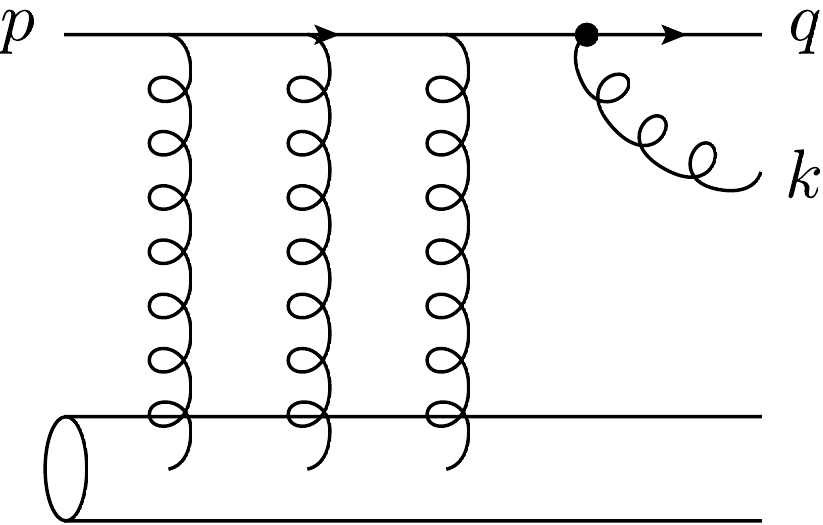}\\{\footnotesize (c)}
 \end{center}
 \end{minipage}
 \caption{(a) Quark production and (b), (c) quark-gluon production in $pA$ collisions.\label{fig:jet}}
 \end{center}
\vspace*{-0.5cm}
 \end{figure}

The propagation of a projectile parton through the CGC is eikonal. With the parton at transverse position $\bmx_{\perp}$ and moving along $x^-$, the interaction with the field $\mcal{A}^{\mu}_a = \delta^{\mu +} \alpha_a$ in covariant gauge is given by the Wilson line
$V^{\dagger}_{\bmx} \equiv {\mbox P}
 \exp\big[ \rmi g\int \rmd x^-\alpha_a(x^-,\bmx_{\perp}) t^a\big]$.
We want to study particle production in $hA$ collisions, with $h$ a dilute projectile hadron and $A$ a dense target, like a large nucleus. The partonic process is $qA\to qX$, cf.~Fig.~\ref{fig:jet}.a, and, in the proton fragmentation region, a large-$x$ quark from the proton interacts with the soft components of the nucleus via the Wilson line. The cross section is
 \beq\label{single}
 \frac{\dif N_q}{\dif y \, \dif^2 \bmp_{\perp}} \sim
 x_1 f_q(x_1,p_{\perp}^2)
 \int \dif^2 \bmr_{\perp}
 \rme^{-\rmi \bmr_{\perp} \cdot \bmp_{\perp}}\,
 \lan \hat{S}_{\bmx_1\bmx_2}\ran_Y. 
 \eeq
Here $\bmp_{\perp}$ and $y$ are the transverse momentum and rapidity of the produced quark, $x_1 f_q$ is the quark distribution in the proton, $\bmr_{\perp} = \bmx_1 - \bmx_2$, $\hat{S}_{\bmx_1\bmx_2} = (1/N_c)\,\rmtr({V}^{\dagger}_{\bmx_1} {V}_{\bmx_2}^{\phantom{\dagger}})$ with the Wilson lines in the fundamental representation is the ``dipole'' $S$-matrix and the average is to be taken with $W_Y[\alpha]$ of the nucleus. The fractions $x_1 = p_{\perp} \rme^{y}/\sqrt{s}$ and $x_2 = \rme^{-Y} = p_{\perp} \rme^{-y}/\sqrt{s} \ll 1$ are determined from the kinematics in the Born approximation. The ratio
 \beq\label{rpa}
 R_{pA} = \frac{1}{A^{1/3}}\,
 \frac{\dif N_h / \dif \eta\, \dif^2 \bmp_{\perp}|_{pA}}
{\dif N_h / \dif \eta\, \dif^2 \bmp_{\perp}|_{pp}}
 \eeq
should be suppressed compared to unity since a big nucleus is more saturated than a proton. $R_{pA}$ is measurable and fits to RHIC data from $dAu$ collisions  are successful and predictions for the $pPb$ ones at LHC are available \cite{Albacete:2012xq}.

Similarly we can study di-hadron correlations in $hA$ collisions in the forward region. One expects a suppression of the azimuthal correlation of the two hadrons when their transverse momenta are of order $\order{Q_s}$, and in fact this has already been observed and explained by CGC at RHIC \cite{Albacete:2010pg,Lappi:2012nh}. Studying the process at the partonic level one can derive the analogous to \eqn{single} cross section for the production of, e.g., a $qg$ pair. It is straightforward to see, from squaring Fig.~\ref{fig:jet}.b, that it is necessary to introduce higher-point correlators involving the product of the quadrupole $\hat{Q}_{\bmx_1\bmx_2\bmx_3\bmx_4} = (1/N_c)\,\rmtr({V}^{\dagger}_{\bmx_1} {V}_{\bmx_2}^{\phantom{\dagger}} {V}^{\dagger}_{\bmx_3} {V}_{\bmx_4}^{\phantom{\dagger}})$ times a dipole. The average is to be done at a large-$Y$ determined by the kinematics of the process.

\section{Multi-gluon correlators in JIMWLK}
\label{sec:tech}

The JIMWLK Hamiltonian determining the QCD evolution of $W_Y[\alpha]$ and hence of the (gauge invariant) correlators is
 \beq\label{H} 
 H = -\frac{1}{16 \pi^3} \int_{\bmu\bmv\bmz}
 \mcal{M}_{\bmu\bmv\bmz}
 \left(1 + \wt{V}^{\dagger}_{\bmu} 
 \wt{V}_{\bmv}^{\phantom{\dagger}}
 -\wt{V}^{\dagger}_{\bmu} 
 \wt{V}_{\bmz}^{\phantom{\dagger}}
 -\wt{V}^{\dagger}_{\bmz} 
 \wt{V}_{\bmv}^{\phantom{\dagger}}\right)^{ab}
 \frac{\delta}{\delta \alpha_{\bmu}^a}\,
 \frac{\delta}{\delta \alpha_{\bmv}^b},
 \eeq
with $\mcal{M}_{\bmu\bmv\bmz} = (\bmu-\bmv)^2/[(\bmu-\bmz)^2(\bmz-\bmv)^2]$ the dipole kernel, the tilde standing for the adjoint representation and the functional derivatives acting on the upper and lower end-points of the Wilson lines $V^{\dagger}$ and $V$ respectively.

\begin{figure}
 \begin{center}
 \begin{minipage}[b]{0.325\textwidth}
 \begin{center}
 \includegraphics[scale=0.5]{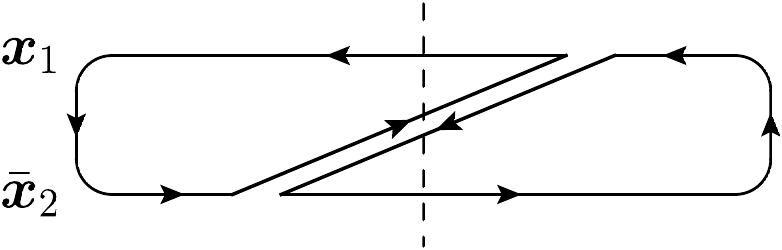}
 \end{center}
 \begin{center}
 \includegraphics[scale=0.5]{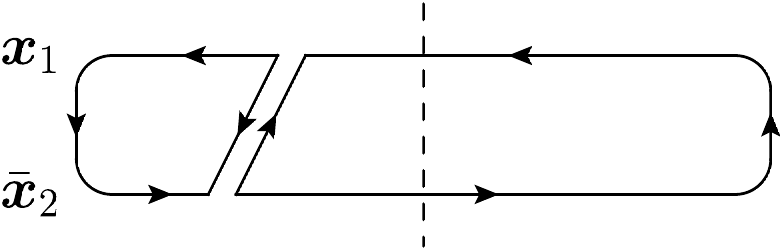}\\{\footnotesize (a)}
 \end{center}
 \end{minipage}
 \begin{minipage}[b]{0.325\textwidth}
 \begin{center}
 \includegraphics[scale=0.5]{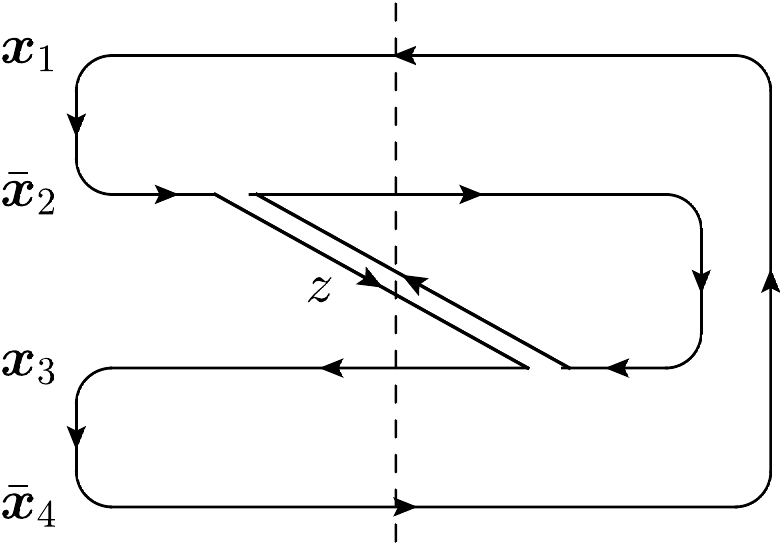} \\{\footnotesize (b)}
 \end{center}
 \end{minipage}
 \begin{minipage}[b]{0.325\textwidth}
 \begin{center}
 \includegraphics[scale=0.5]{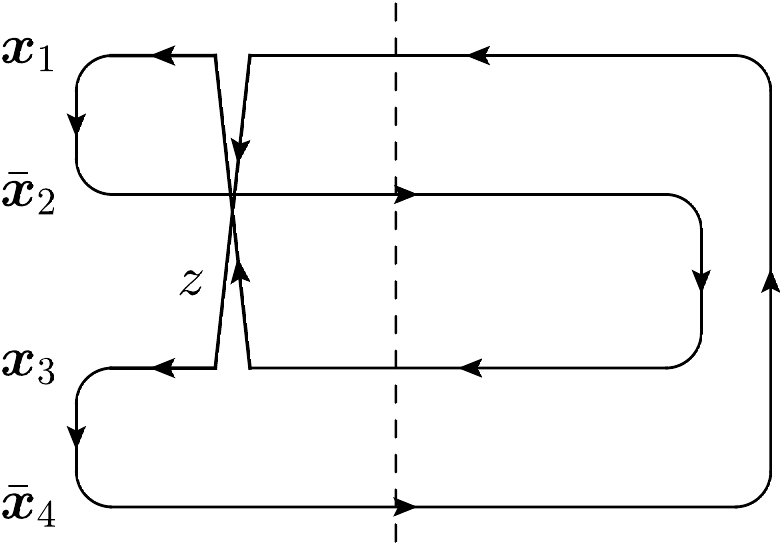}\\{\footnotesize (c)}
 \end{center}
 \end{minipage}
 \caption{(a) Real and virtual emission of a $q\bar{q}$ pair from a dipole. (b) and (c) Similarly from a quadrupole. The dashed line denotes the interaction with the target.
 \label{fig:dipquad}}
 \end{center}
 \vspace*{-0.5cm}
 \end{figure}

One method to calculate the correlators is to reformulate \eqn{JIMWLK} into a Langevin equation \cite{Blaizot:2002xy} for a Wilson line and solve on a lattice. The other is to construct the evolution equations for the correlators $\lan \hat{\mathcal{O}} \ran_Y$ of interest using Eqs.~\eqref{JIMWLK} and \eqref{H}. The structure of the real terms (the last two) of the Hamiltonian is such that it leads to a hierarchy of equations \cite{Balitsky:1995ub}; $H_{\rm real}$ acting on $n$ Wilson lines, can lead to $n+2$ of them. This second method is sufficient, if, for instance, one wants to calculate single inclusive gluon production in $pA$ collisions, cf.~\eqn{single}. JIMWLK leads to the dipole equation
 \beq\label{BK}
 \frac{\del \lan \hat{S}_{\bmx_1\bmx_2} \ran_Y}{\del Y}=
 \frac{\abar}{2\pi}\, \int_{\bmz}
 \mcal{M}_{\bmx_1\bmx_2\bmz}
 \lan \hat{S}_{\bmx_1\bmz} \hat{S}_{\bmz\bmx_2}
 -\hat{S}_{\bmx_1\bmx_2} \ran_Y,
 \eeq
which is interpreted in terms of projectile evolution as shown in Fig.~\ref{fig:dipquad}.a at large-$N_c$. One factorizes $\lan \hat{S} \hat{S}\ran_Y = \lan \hat{S} \ran_Y \lan \hat{S} \ran_Y$ to get the BK equation \cite{Balitsky:1995ub,Kovchegov:1999yj}, a closed equation whose solution is known semi-analytically and numerically.

Less inclusive quantities require the knowledge of higher-point correlators. The quadrupole equation \cite{JalilianMarian:2004da} is more involved and two representative diagrams are shown in Figs.~\ref{fig:dipquad}.b and \ref{fig:dipquad}.c. Even at large-$N_c$, where it becomes a closed inhomogeneous and linear equation (requiring the input of $\lan \hat{S} \ran_Y$), the large number of transverse variables and the non-locality in transverse space seem to be prohibitive for the possibility of a numerical solution.
 
The implementation of the Langevin approach \cite{Dumitru:2011vk}, lead to the finding that the numerical data for the quadrupole are very well-described by a Gaussian mean field approximation (MFA); an extrapolation to arbitrary $Y$ of the McLerran-Venugopalan (MV) model, the typical initial condition. Such a Gaussian $W_Y[\alpha]$ involves a single kernel, hence all high-point correlators are expressed in terms of the 2-point one, although there was no a priori reason for this to happen for the, highly non-linear, JIMWLK evolution. 

Still, one can prove that a Gaussian approximation is a quasi-exact solution to the JIMWLK equation \cite{Iancu:2011ns,Iancu:2011nj}. At saturation, real emissions are suppressed and the virtual part $H_{\rm virt}$ (the first two terms) of the Hamiltonian dominates. This is Gaussian, including the second term; those Wilson lines transform the ``left'' derivatives to ``right'' ones which act on the lower and upper end-points of the Wilson lines $V^{\dagger}$ and $V$ respectively. We integrate the kernel over $\bmz$ for $1/Q_s \ll |\bmu-\bmz|,|\bmv-\bmz| \ll |\bmu-\bmv|$, with the lower limit imposed by our approximation, while the upper gives the dominant logarithmic contribution $2 \ln[(\bmu-\bmv)^2 Q_s^2]$. This would lead to a result valid only at saturation, but doing the same approximation in \eqn{BK} we find this logarithm to be related to the logarithmic derivative of the dipole. Thus
 \beq\label{HG}
 H_{\rm G} = \frac{1}{4 g^2 C_F}
 \int_{\bmu\bmv}
 \frac{\dif \ln \lan \hat{S}_{\bmu\bmv}\ran_Y}{\dif Y}\,
 \left(\frac{\delta}{\delta \alpha_{{\rm L}\bmu}^a} 
 \frac{\delta}{\delta \alpha_{{\rm L}\bmv}^a}
 +
 \frac{\delta}{\delta \alpha_{{\rm R}\bmu}^a} 
 \frac{\delta}{\delta \alpha_{{\rm R}\bmv}^a}
 \right).
 \eeq
This is a Gaussian Hamiltonian and the kernel can be most easily determined from the BK equation. \eqn{HG} is correct, at finite-$N_c$, at saturation by construction and in the dilute limit as can be inspected.

$H_{\rm G}$ generates evolution equations for multi-gluon correlators which are local in the transverse plane, that is, they are ordinary differential equations in $Y$ with $Y$-dependent coefficients. A further ``separability'' property of the kernel in \eqn{HG} leads to correlators which are local function in $Y$. For instance, the quadrupole at large-$N_c$ with MV model initial condition reads
 \begin{align}\label{Qsol}
 \lan \hat{Q}_{1234} \ran_Y
 = &\frac{\ln\big[\lan \hat{S}_{12}\ran_Y
 \lan \hat{S}_{34}\ran_Y
 /\lan \hat{S}_{13}\ran_Y
 \lan \hat{S}_{24}\ran_Y\big]}
 {\ln\big[\lan\hat{S}_{12}\ran_Y
 \lan \hat{S}_{34}\ran_Y
 /\lan \hat{S}_{14}\ran_Y
 \lan \hat{S}_{23}\ran_Y\big]}\, 
 \lan \hat{S}_{12}\ran_Y
 \lan \hat{S}_{34}\ran_Y
\nn
 +
 &\frac{\ln\big[\lan \hat{S}_{14}\ran_Y
 \lan \hat{S}_{23}\ran_Y
 /\lan \hat{S}_{13}\ran_Y
 \lan \hat{S}_{24}\ran_Y\big]}
 {\ln\big[\lan\hat{S}_{14}\ran_Y
 \lan \hat{S}_{23}\ran_Y
 /\lan \hat{S}_{12}\ran_Y
 \lan \hat{S}_{34}\ran_Y\big]}\,  
 \lan \hat{S}_{14}\ran_Y
 \lan \hat{S}_{23}\ran_Y
 \end{align}
(with $i=\bmx_i$). It obeys a mirror symmetry $\lan\hat{Q}_{1234} \ran_Y = \lan\hat{Q}_{1432} \ran_Y$, which holds at finite-$N_c$ and beyond the MFA and is a result of time-reversal symmetry, where time stands for $x^{-}$ \cite{Iancu:2011nj}. It is preserved by JIMWLK due to the presence of both left and right derivatives and suggests that the hadron expands symmetrically in the $x^{-}$ direction. At the level of the MFA only, it is also symmetric under the charge conjugation $\lan\hat{Q}_{1234} \ran_Y = \lan\hat{Q}_{2341} \ran_Y$.

\begin{figure}[t]
 \begin{center}
 \begin{minipage}[b]{0.45\textwidth}
 \begin{center}
 \includegraphics[scale=0.4]{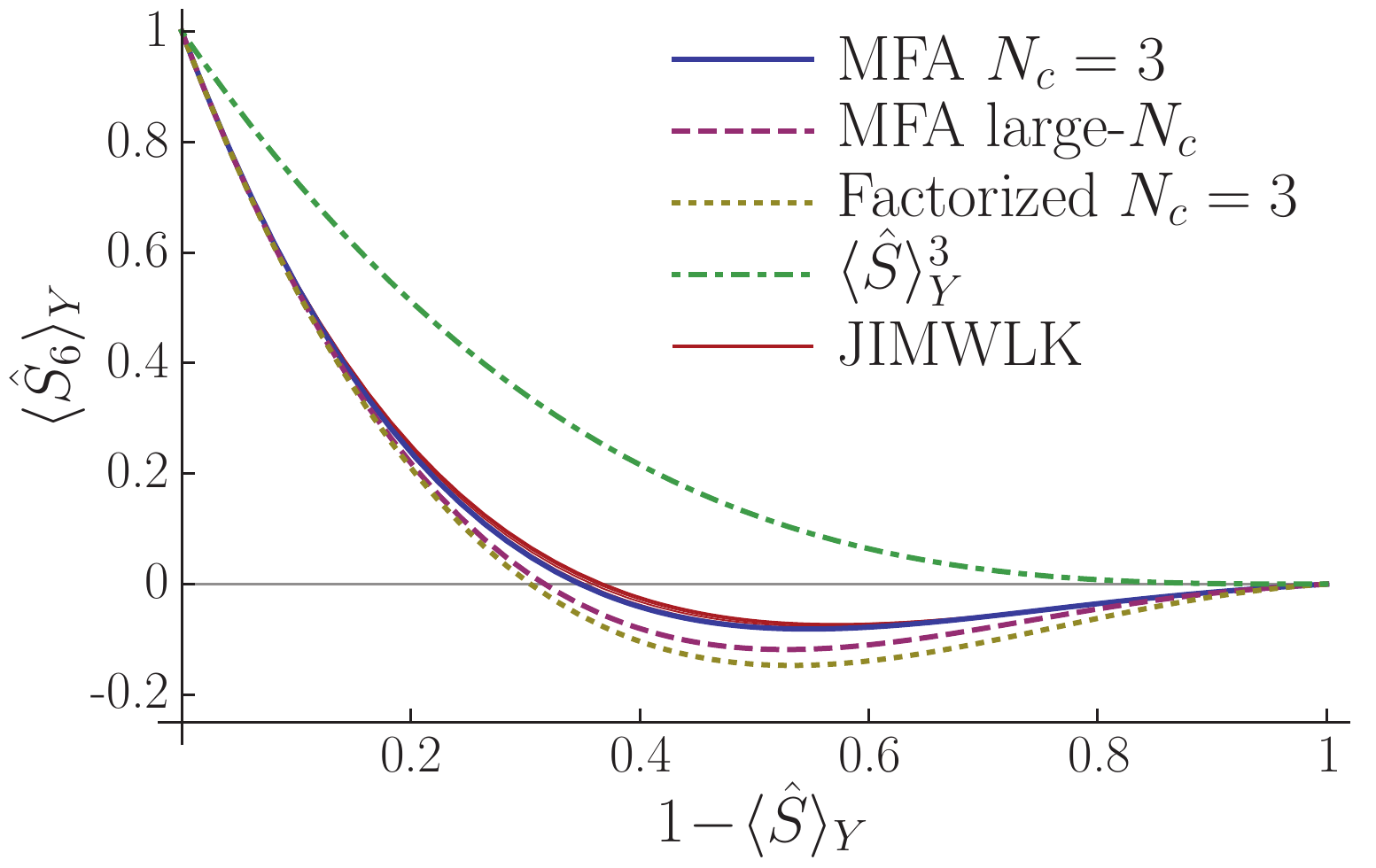}\\{\footnotesize (a)}
 \end{center}
 \end{minipage}
 \qquad
 \begin{minipage}[b]{0.45\textwidth}
 \begin{center}
 \includegraphics[scale=0.8]{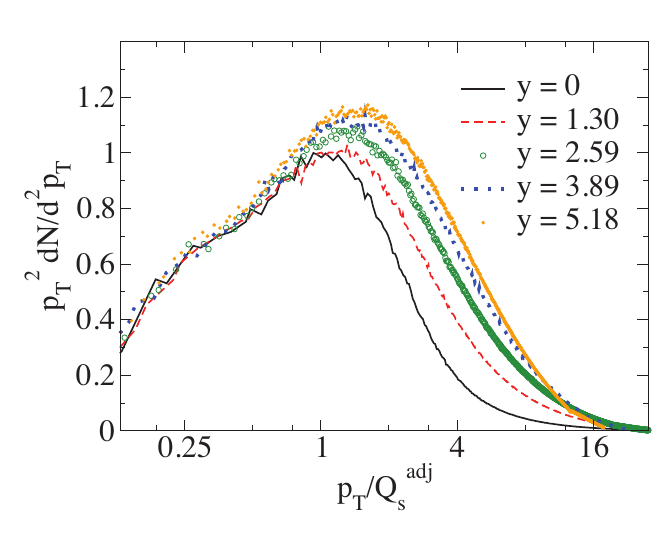}\\{\footnotesize (b)}
 \end{center}
 \end{minipage}
 \caption{(a) $\lan \hat{S}_6 \ran_Y = (N_c^2/(N_c^2-1)
 \lan \hat{Q}_{1234} 
 \hat{S}_{41} - (1/N_c^2)\hat{S}_{32}\ran_Y$ with $\bmx_3=\bmx_1$ and $\bmx_4=\bmx_2$. Red: JIMWLK from $Y=0$ to $5.18$ \cite{Dumitru:2011vk}. Blue: MFA analytic result for $N_c=3$ \cite{Iancu:2011nj}. At $Y=0$ JIMWLK and MFA coincide because of the initial condition. True at any $Y$ in the dilute and in the dense region. A tiny difference, stabilizing soon, occurs in the transition region. (b) Scaling of the gluon spectrum\cite{Lappi:2011ju}.\label{fig:S6gluon}}
 \end{center}
 \vspace*{-0.5cm}
 \end{figure}

At finite-$N_c$ operators mix, e.g.~the phenomenologically interesting 6-point operator $\hat{Q}\hat{S}$ mixes with two more and the equation from the matrix diagonalization is cubic. As shown in Fig.~\ref{fig:S6gluon}.a for a simple configuration the agreement of the analytic and numerical results is excellent. Analytic expressions for multi-gluon correlators are necessary to reduce the numerical cost for Fourier transforming expressions like \eqn{single} (and more complicated ones, like in the double inclusive case) and obtain the desired cross sections.

\section{More properties and applications}
\label{sec:more}

One expects quantities to scale with $Q_s$ \cite{Iancu:2002tr,Mueller:2002zm}. This property, first observed in DIS \cite{Stasto:2000er}, extends to other quantities, like gluon production at early times in $AA$ collisions \cite{Lappi:2011ju}, cf.~Fig.~\ref{fig:S6gluon}.b. In such collisions, due to final state interactions, the CGC cannot describe spectra of produced hadrons, but it can give more inclusive quantities, e.g.~total multiplicities \cite{Iancu:2012xa}.

\providecommand{\href}[2]{#2}
\begingroup\raggedright

\endgroup


\end{document}